\documentclass
[journal,10pt,draftclsnofoot,onecolumn]{IEEEtran}

\IEEEoverridecommandlockouts
\usepackage{ifpdf}
\usepackage{cite}
\ifCLASSINFOpdf
   \usepackage[pdftex]{graphicx}
 \DeclareGraphicsExtensions{.eps}
\else
   \usepackage[dvips]{graphicx}
  \DeclareGraphicsExtensions{.eps}
\fi
\usepackage[cmex10]{amsmath}
\usepackage{algorithmic}
\usepackage{array}
\usepackage{mdwmath}
\usepackage{mdwtab}
\usepackage{eqparbox}
\usepackage{algorithm}
\usepackage{algorithmic}

\usepackage{url}
\usepackage{mdwmath}
\usepackage{mdwtab}
\usepackage{amsmath}

\usepackage{amssymb}
\usepackage{dsfont}
\begin{document}
\title{To Relay or Not To Relay in Cognitive Radio Sensor Networks}
\author{\IEEEauthorblockN{F.~Foukalas$^{*}$, \textit{Member, IEEE}, T.~Khattab$^{*}$, \textit{Member, IEEE\\}}
\IEEEauthorblockA{$^*$Electrical Engineering, Qatar University, Doha, Qatar\\}}

\maketitle

\begin{abstract}
Recent works proposed the relaying at the MAC layer in cognitive radio networks whereby the primary packets are forwarded by the secondary node maintaining an extra queue devoted to the relaying function. However, relaying of primary packets may introduce delays on the secondary packets (called secondary delay) and require additional power budget in order to forward the primary packets that is especially crucial when the network is deployed using sensors with limited power resources. To this end, an admission control can be employed in order to manage efficiently the relaying in cognitive radio sensor networks. In this paper, we first analyse and formulate the secondary delay and the required power budget of the secondary sensor node in relation with the acceptance factor that indicates whether the primary packets are allowed to be forwarded or not. Having defined the above, we present the tradeoff between the secondary delay and the required power budget when the acceptance factor is adapted. In the sequel, we formulate an optimization problem to minimize the secondary delay over the admission control parameter subject to a limit on the required power budget plus the constraints related to the stabilities of the individual queues due to their interdependencies observed by the analysis. The solution of this problem is provided using iterative decomposition methods i.e. dual and primal decompositions using Lagrange multipliers that simplifies the original complicated problem resulting in a final equivalent dual problem that includes the initial Karush Kuhn Tucker conditions. Using the derived equivalent dual problem, we obtain the optimal acceptance factor while in addition we highlight the possibilities for extra delay minimization that is provided by relaxing the initial constraints through changing the values of the Lagrange multipliers. Finally, we present the behaviour of secondary delay in case infinite and finite queues are employed and thus the overflow and blocking probabilities are assessed respectively.        
\end{abstract}

\begin{keywords}
packet relaying, queues, cognitive radio sensor networks, optimization, Lagrange duality, decomposition methods.   
\end{keywords}

\IEEEpeerreviewmaketitle
\section{Introduction}
Cognitive radio (CR) has attracted a lot of interest in the last decade from research and industrial communities of communications and networking. In general, a CR system allows the spectrum utilization of a licensed frequency channel by a secondary system (unlicensed) without significantly  affecting the efficiency of spectrum utilization of the primary (licensed) user ~\cite{c1}. In order to exploit this functionality, a new type of deployment is needed using nodes that acquire the knowledge of channels’ occupancy. The new demands for CR networks (CRNs) deployment have shown via the standardization activities that a new type of wireless sensor networks (WSNs) will be shaped targeting to exploit the benefits of the efficient spectrum utilization ~\cite{c2}. This type of CRNs were named cognitive radio sensor networks (CRSNs) that require a customized handling in order to achieve the desire results ~\cite{c3}.     

Several issues are raised from such a new application of WSNs such as the error free spectrum sensing (SpSe), new communication protocols based on the SpSe results ~\cite{c4} and new resource allocation mechanisms as packet relaying through the secondary nodes ~\cite{c5}. Packet relaying has originally proposed first in ~\cite{c6} in which a secondary node retains two queues one for its own secondary packets and one for providing the relay of primary packets. Based on this concept, the authors studied the maximum allowable throughput achieved by the cognitive user while maintaining the stability of the overall distributed queueing system. However, the delay imposed on the secondary packets was not studied. Moreover, the additional power budget required for the relaying process were not taken into account that is very important in case of CRSNs with limited power resources. A recent work in ~\cite{c7} has considered a cognitive relaying framework that manage to forward relaying information; however, although delay expressions were given, still power constraints were not considered and an optimization solution is not provided in general. A more recent work in ~\cite{c8} presents the tradeoff between the sensing and the energy consumption with relays; however, the relaying process is not taken place at the MAC layer but at the physical layer looking into corresponding cooperative strategies. In ~\cite{c9}, authors first describe the delay and the power consumption required by the relaying process in CRSNs and in this paper we extend this work defining and solving a specific optimization problem highlighting useful insights in this new topic. 

Specifically, we analyse the delay of secondary packets and the power consumption which are both requirements for the implementation of the packet relaying at the secondary node. We consider that the secondary node (i.e. cognitive) retains its own traffic in the secondary queue while the relayed primary packets are stored in a separate devoted primary queue. The secondary node is able to control the admission of primary packets passed through the relaying primary queue using an admission control parameter. We assume a scheduling strategy that provides priority to relayed primary traffic over the secondary one. Based on this cognitive radio model, we obtain the formulation of the secondary delay and the required power budget whereby we highlight the following relaying trade-off: \textit{on one hand, the admission of more primary packets will reduce the secondary delay and on the other hand, it will increase the required power budget}. This particular tradeoff can be investigated taking into account different objectives. In our case, we opt to formulate a minimization problem for the secondary delay over the admission control parameter i.e. acceptance factor subject to the constraint in the power budget and the separate stabilities' rules of each individual queue. Using optimization theory and decomposition methods in particular, since the original is complicated and depends on several constraints, we prove that indeed an optimal acceptance factor value exists under several conditions in which the minimum delay is achieved and yet the power budget is sufficiently kept in the specific upper bound. Finally, we discuss the relaying behaviour considering different queue models for the secondary queue in terms of buffer size (i.e. infinite or finite) giving more light from practical point of view.

The rest of this paper is organized as follows: Section II provides the system model and the necessary assumptions. The analysis is provided in section III, and in section IV we formulate the optimization problem and in the sequel we provide its solution. Section V presents the different behaviour of the relaying queue with infinite and finite buffers while the simulation results are presented in section VI. This paper is concluded with section VII.

\section{System Model} \label{system}
We consider the cognitive radio sensor network depicted in Fig. 1 that consists of a pair of a primary user (PU) transmitter and receiver denoted as PU-Tx and PU-Rx respectively and a pair of a secondary transmitter (SU-Tx) and receiver (SU-Rx). There are three links of interest namely the primary link between the PU-Tx and PU-Rx with a channel gain denoted by $g_p$, the link between the SU-Tx and PU-Tx with a channel gain denoted by $g_{s,p}$ and the link between the PU-Rx and SU-Tx with a channel gain denoted by $g_{p,s}$. All channel gains are assumed to be zero mean circularly symmetric complex Gaussian random variable with variance $\sigma_i^2$ and independent for all $i$ where $i$ takes the values $'p'$, $'s'$, $'ps'$ and $'sp'$ indicating the primary, secondary, primary to secondary and secondary to primary links respectively. Each link is affected by complex additive white Gaussian noise AWGN with zero mean and unit variance and independent for all links.  We assume that successful transmission is achieved when the instantaneous signal to noise ratio (SNR) on the $i-th$ link, employing the maximum transmission power $P_i$, given by $\mid g_i \mid^2 P_i$, exceeds a predefined threshold $\gamma_{th,i}$.

The considered scenario throughout this paper is that the SU-Tx utilizes the  spectrum resource in order either to serve its secondary receiver (SU-Rx) whenever the primary link is idle or  to serve the PU-Rx by relaying the primary packets in case the primary link is not able to achieve adequate communication (i.e. outage condition). In this fashion, the SU-Tx plays two roles in the presented CR scenario such as giving on one hand the opportunity to forward the primary packets whenever the primary link experiences outage conditions and on the other hand to exploit the spectrum for itself whenever is not used by the primary link thus forwarding the secondary packets to its own user. Henceforth, we consider that the SU-Tx maintains two queues denoted as $Q_s$  and $Q_{ps}$  for serving its own packer i.e. secondary packets and the primary packets respectively. Moreover, we assume that the PU-Tx retains a queue $Q_p$ to store its own primary packets. Afterwards, the presented model assumes that the SU-Tx has the capability to relay or not to relay the primary packets and this is a matter of making a decision taking into account the current channel conditions. 

\begin{figure}
  \includegraphics[width=\columnwidth]{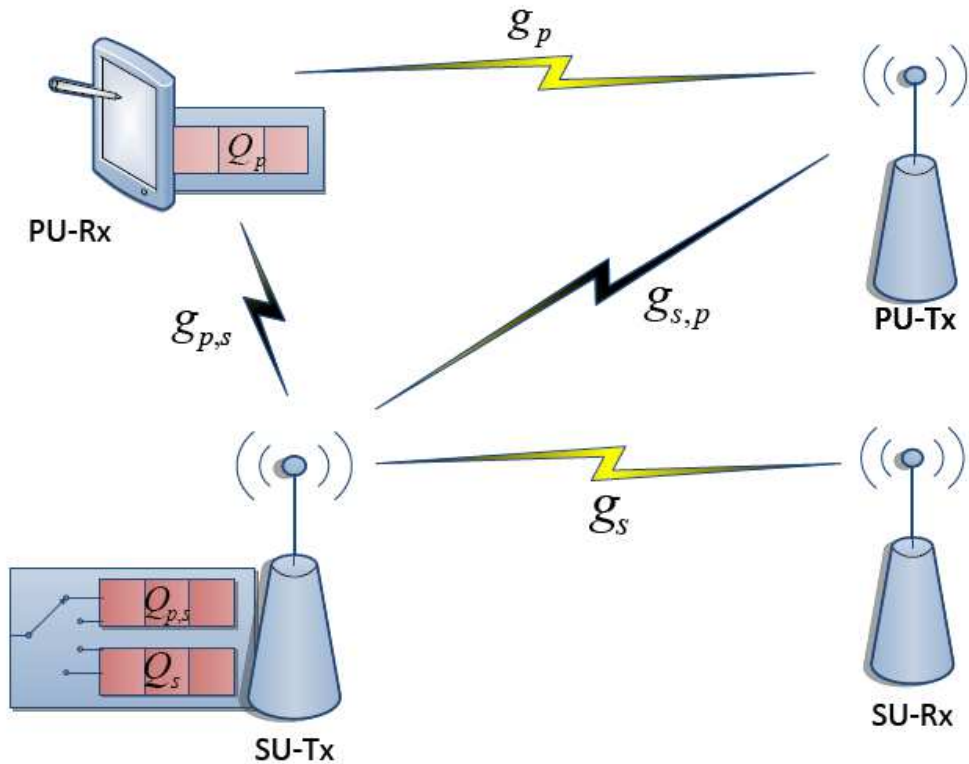}\\
  \caption{System model of cognitive radio sensor network with packet relaying capabilities}
  \label{fig:systemmodel}
\end{figure}

For simplicity we assume that all packets are transmitted in a time-slotted fashion where they have the same size and one time slot is required for transmission of a single packet. Initially, we assume that all queues are of infinite lengths. The packet arrival processes of self-traffic at each node are independent and with mean arrival rates $\lambda_p$ and $\lambda_s$ (packet/slot) for the $Q_p$ and $Q_s$ queues respectively. The MAC layer is assumed to obey the following protocol. At a certain time slot and given the priority to grant the PU unconditional access to the channel, the PU-Tx transmits a packet to the primary destination i.e. PU-Rx. A correctly received packet by the primary destination is acknowledged by sending an ACK message resulting in the primary queue $Q_p$ dropping this packet. In case of packet failure (indicated by ACK timeout) and assuming that the SU-Tx can listen to all PU-Tx transmissions, the SU-Tx can accept a fraction of the undelivered primary packets in its relaying queue $Q_{ps}$, if it was able to correctly decode them and send ACK messages to the primary destination PU-Rx. In case both the SU-Tx and the PU-Rx fail to decode the primary data, a retransmission of the packet is initiated by the PU-Tx. We assume that the overhead for transmitting the ACK and NACK messages is very small compared to packet sizes. In addition, we assume that the ACK messages are always  decoded perfectly at the PU-Rx and the SU-Tx. At the beginning of each time slot, the SU-Tx is allowed to sense the channel and if it is declared idle, the SU-Tx transmits from one of its queues under the conditions discussed below.

\section{Performance Analysis and Metrics of Interest} \label{system}
In this section, we present, the analysis and the corresponding metrics of interest for the considered cognitive sensor node with packet relaying capabilities. We tackle with two performance metrics and obtain their formulation and how these metrics are involved in the decision making mechanism at the SU-Tx for deciding on relaying the primary packets. The metrics considered are the SU-Tx's delay introduced to the secondary packets in the $Q_s$, called secondary delay and the power budget required by the $Q_{ps}$ to forward the primary packets. 

\subsection{Relaying Factor and Scheduling} \label{system}
We assume a relaying factor $f \in [0,1]$ that is managed by the SU-Tx for making the decision to relay or not relay the secondary packets. Increasing $f$ to $1$ means that no admission control is accomplished at the $Q_{ps}$ thus relaying with any restriction is taken place and making $f$ equal to $0$ means that no relaying is accomplished thus the packets are blocked from the assumed admission control mechanism. Relied on this factor, the SU-Tx performs the following scheduling strategy called Prioritized Cognitive Relaying (PCR) where the primary packets have always priority against the secondary ones ~\cite{c10}. In particular, using this strategy, when an idle slot is detected, the SU-Tx attempts to access the primary channel in order to transmit first the primary packets using the relaying queue $Q_{ps}$. If possible, the SU-Tx uses the minimum transmission power level sufficient to achieve the target (Signal-to-Noise-Ratio) SNR threshold $\gamma_{th,sp}$  at the primary destination which is equal to $\gamma_{th,sp}/ \mid g_{sp} \mid ^2$. Thus, an outage probability is defined for each link denoted as $P_{out,i}$ with $i$ the index for each link. If the link between the relay and the primary destination experiences a deep fade or the relaying queue is empty, the SU-Tx will transmit from the secondary traffic queue $Q_s$ (its own traffic) with the same procedure. According to the adopted PCR, the arrival and service rates of the CRSN denoted as $\lambda_i$ and $\mu_i$ for each $i$ queue respectively encompass the relaying factor $f$ in their formulation as follows:

\begin{itemize}
\item[-] Primary Queue Service Rate $\mu_p$: The primary queue $Q_p$ will forward the packets once the primary link does not experience outage and when the primary to secondary link does not also experience outage taking into account the relaying factor and thus it is obtained as   
\begin{eqnarray} \label{eq1}
 \mu_p = (1-P_{out,p}) + f P_{out,p}(1-P_{out,ps}) 
\end{eqnarray}
\item[-] Relaying Queue Arrival Rate $\lambda_{ps}$: The relaying queue $Q_{ps}$ will arrive the packets once the primary link does not experience outage taking into account the relaying factor and when the primary to secondary link does not experience outage taking into account the outage of the primary link  and thus it is obtained as 
\begin{eqnarray} \label{eq2}
 \lambda_{ps} = f P_{out,p} + f P_{out,p}(1-P_{out,ps})\frac{\lambda_p}{\mu_p} 
\end{eqnarray}
\item[-] Relaying Queue Service Rate $\mu_{ps}$: The relaying queue $Q_{ps}$ will forward the packets once the secondary to primary link does not experience outage and thus it is obtained as   
\begin{eqnarray} \label{eq3}
 \mu_{ps} = (1-\frac{\lambda_p}{\mu_p})(1-P_{out,ps}) 
\end{eqnarray}
\item[-] Secondary Queue Service Rate $\mu_s$: The secondary queue $Q_s$ will forward the secondary packets once the secondary to primary link does not experience outage and when the secondary link is also free of outage and thus it is obtained as 
\begin{eqnarray} \label{eq4}
 \mu_{s} = (1-\frac{\lambda_p}{\mu_p})(1-P_{out,s}) (1-\frac{\lambda_{ps}}{\mu_{ps}}(1-P_{out,sp}))
\end{eqnarray}  
\end{itemize}
where the outage probability $P_{out,i}$ for a given threshold in SNR $\gamma_{th,sp}$ is obtained as follows
\begin{eqnarray} \label{eq5}
  P_{out,i}=Pr[\mid g_{sp} \mid^2 P_i<g_{th,i}]=1-e^\frac{\gamma_{th,i}}{\sigma_i^2 P_i}
\end{eqnarray}  

\subsection{Queuing Modeling and Secondary Delay} \label{system}
In this section, we analyse the secondary delay i.e. the delay experienced at the queue $Q_s$ based on the considered  traffic models for all queues in the system. Firstly, we will deal with the primary queue $Q_p$. Per the discussion given in ~\cite{c11}‎ and ~\cite{c12} about the dominant systems, it is concluded that the primary queue can be uncoupled from the overall system of queues. This isolation allows us to model the primary queue as an $M/M/1$ queue. Hence, the expected value of the delay imposed on the primary packets in the primary queue is given by:
\begin{eqnarray} \label{eq6}
 \bar{D}_p = \frac{1-\lambda_p}{\mu_p-\lambda_p}
\end{eqnarray}

Notably, each node chooses its arrival rate based on the application such that satisfying the stability conditions of its queue. For more detailed discussion about the stability and its conditions, the reader is referred to ~\cite{c12}. For the relaying queue $Q_{ps}$ and the secondary queue $Q_s$, we model their traffic model by adopting an $M/G/1$ priority queuing model. While their inter-arrival times are Poisson random variables, we can model their service times ($S_{ps}$ and $S_s$) as Geometric random variables based on the description of the scheduling strategy in ‎~\cite{c10}. The Geometric random variables parameters $p_{ps}$  and $p_s$ are given by $\mu_{ps}$ and $\mu_s$ for the relaying queue and the secondary queue respectively. The first and the second moments of the average service time for the relaying queue ($E[S_{ps}]$ and $E[S_{ps}^2]$) are given as follows
\begin{eqnarray} \label{eq7}
 E[S_{ps}] = \frac{1}{p_{ps}}
\end{eqnarray}
\begin{eqnarray} \label{eq8}
 E[S_{ps}^2] = \frac{1-p_{ps}}{p_{ps}^2}
\end{eqnarray}
With the same logic, we can obtain the first and the second moments of the average service time for the secondary queue i.e. $E[S_s]$ and $E[S_s^2]$. 

We are now ready to analyse and derive the secondary delay based on the delay experienced at secondary packets i.e. secondary queue $Q_s$ that can be produced by three distinguished factors according to the model presented in ‎~\cite{c13}. Taking also into account the scheduling strategy presented above, the different delay factors for a secondary packet of interest inside the secondary queue is assembled as follows: 

\begin{enumerate}
\item The waiting time for the currently transmitted packet being served in the secondary queue $Q_s$ i.e. once the packet of interest is arrived well known as the ‘residual’ time denoted as $\bar{D_1}$
\item The waiting time for being served the packets already appeared in the secondary queue $Q_s$ upon the arrival of the packet of interest denoted as $\bar{D_2}$    
\item The waiting time for being served the primary packets either from the primary queue $Q_p$ or the relaying queue $Q_{ps}$ that will be appeared in that queues after the arrival of the packet of interest in the secondary queue $Q_s$ denoted as  $\bar{D_3}$      
\end{enumerate}
We donate as $\rho_i$ the utilization factor for each queue $i$ that is given in general by the formula $\rho = \lambda_i E[S_i]$. The expected value of the delay $\bar{D_1}$ will be the sum of the residual time at the secondary queue and the primary queue obtained as follows 
\begin{eqnarray} \label{eq9}
  \bar{D_1} = \frac{\rho_s}{1-\rho_p}\frac{E[S_s^2]}{2E[S_s]}+\frac{\rho_ps}{1-\rho_p}\frac{E[S_{ps}^2]}{2E[S_{ps}]}
\end{eqnarray}
where $\rho_i/(1-\rho_p)$ is the conditional probability, that the packet of $i-th$ user is being transmitted given that no primary packet is transmitted. We consider now the second component delay $\bar{D_2}$  which as explained is due to other packets in the secondary queue found by the new arrival. Considering the \textit{ Little’s} law of queuing theory on which the long-term average number of customers in a stable system is equal to the long-term average effective arrival rate, $\lambda$, multiplied by the average time a customer spends in the system, we obtain the following formulation considering the summary of the waiting times introduced from the primary and secondary queues respectively 
\begin{eqnarray} \label{eq10}
  \bar{D_2} = \rho_p \bar{D_p} + \rho_s \bar{D_s}
\end{eqnarray}
In the same concept i.e. the Little’s law, we can define the third component delay $\bar{D_3}$ where only the delay from the primary packets is considered as follows
\begin{eqnarray} \label{eq11}
  \bar{D_3} = \rho_p \bar{D_p} 
\end{eqnarray}
By adding the three components we can obtain the delay of the secondary queue $Q_s$ that eventually become  
\begin{eqnarray} \label{eq12}
  \bar{D_s} = \frac{\frac{\rho_s}{1-\rho_p}\frac{E[S_s^2]}{2E[S_s]}+\frac{\rho_{ps}}{1-\rho_p}\frac{E[S_{ps}^2]}{2E[S_{ps}]}+2\rho_p \bar{D_p}}{1-\rho_s}
\end{eqnarray}
\subsection{Relaying Power Budget} \label{system}
Except the delay at the secondary queue $\bar{D_s}$, the other important metric for our model is the power consumed on the relayed traffic relative to the total power consumed by the SU-Tx. An expression for the average power consumed by the SU-Tx on its own traffic as well as the relayed packets is analysed below. To this end, we define the relaying power budget as \textit{the ratio between the average power per slot consumed on the relaying effort $P_{relay}$ and the total average power per slot consumed $P_{total}$} that is expressed as follows
\begin{eqnarray} \label{eq13}
  \Gamma = \frac{P_{relay}}{P_{total}} = \frac{\lambda_{ps}E[P_{sp}]}{\lambda_{ps}E[P_{sp}]+\lambda_{s}E[P_{s}]}
\end{eqnarray}
where $E[P_{sp}]$ is the average transmission power per relayed packet that is obtained as below considering the assumption of the considered CRSN. First, the transmission power consumed on a relayed primary packet is 
\begin{eqnarray} \label{eq14}
  P_{sp} = \frac{\gamma_{th,sp}}{\mid g_{sp} \mid^2}
\end{eqnarray}
where it is a random variable since it depends on the channel gain $\mid g_{sp} \mid^2$ which is a chi-square distributed random variable. To calculate the average power per packet consumed on relaying, we need the probability density function (PDF) $P_{sp}$ that is defined as follows
\begin{eqnarray} \label{eq15}
  p(x) = \frac{1}{\sigma_{sp}^2} e^{\frac{x}{\sigma_{sp}^2}}, & & \forall x \geq 0
\end{eqnarray}
where $x=\mid g_{sp} \mid^2$ that is obviously an exponential random variable. Considering both (\ref{eq13}) and (\ref{eq14}), we can assume the following for the PDF
\begin{eqnarray} \label{eq16}
  p(x) = \frac{\gamma_{th,sp}}{\sigma_{sp}^2 P_{sp}^2} e^{\frac{\gamma_{th,sp}}{\sigma_{sp}^2 P_{sp}^2}}, & & \forall P_{sp} \geq 0
\end{eqnarray}
Since the $P_{sp}$ cannot exceed the maximum limit $P_{sp}^*$ then the average power can be obtained as follows 
\begin{eqnarray} \label{eq17}
\nonumber
  & & E[P_{sp} \leq P_{sp}^*] = p(x) = \int_0^{P_{sp}^*} P{sp}p(P_{sp}) dP_{sp} \\
  &=& \int_0^{P_{sp}^*} \frac{\gamma_{th,sp}}{\sigma_{sp}^2 P_{sp}} e^{-\frac{\gamma_{th,sp}}{\sigma_{sp}^2 P_{sp}^2}}  dP_{sp}, \forall P_{sp} \geq 0
\end{eqnarray}
Let $z=\frac{\gamma_{th,sp}}{\sigma_{sp}^2 P_{sp}}$ and then $dP_{sp} \frac{\gamma_{th,sp}}{\sigma_{sp}^2} \frac{-dz}{z^2}$ that will give the lower limit of the integration as equal to $\frac{\gamma_{th,sp}}{\sigma_{sp}^2 P_{sp}}$ and the upper limit as infinite where finally we have 
\begin{eqnarray} \label{eq18}
 E[P_{sp}]= \frac{\gamma_{th,sp}}{\sigma_{sp}^2} \int_{ \frac{\gamma_{th,sp}}{\sigma_{sp}^2 P_{sp}}}^{\infty} z^{-1} e^{-z} dz
\end{eqnarray}
The integral in (\ref{eq18}) is known as the upper incomplete gamma function and is calculated using ‘The Exponential Integral’ ~\cite{c13} denoted as $E_1[.]$ and thus we have the following formulation 
\begin{eqnarray} \label{eq19}
 E[P_{sp}]= \frac{\gamma_{th,sp}}{\sigma_{sp}^2} E_1(\frac{\gamma_{th,sp}}{\sigma_{sp}^2 P_{sp}}) = \frac{\gamma_{th,sp}}{\sigma_{sp}^2} \varepsilon_{sp}
\end{eqnarray}
Substituting (\ref{eq19}) into (\ref{eq13}) and maintaining some simplification, yields
\begin{eqnarray} \label{eq20}
 \Gamma = \frac{\lambda_{ps} \gamma_{th,sp} \varepsilon_{sp} \sigma_s^2}{\lambda_{ps} \gamma_{th,sp} \varepsilon_{sp} \sigma_s^2 + \lambda_s \gamma_{th,s} \varepsilon_{s} \sigma_{sp}^2}
\end{eqnarray}

\section{Relaying Tradeoff at MAC Layer of Secondary Node}
In the previous section, we analysed and established the formulation of the secondary delay $\bar{D_s}$ and the required power budget $\Gamma$ of the secondary queue $Q_s$ in relation to the acceptance factor $f$ of the admission control policy applied at $Q_s$. In this section, we present the relaying tradeoff by which the decision making to relay or not to relay will be built. This decision is based on the fact that on one hand increasing the acceptance factor $f$ the average delay $\bar{D_s}$ increases and on the other hand the power budget increased is proportional to acceptance factor as depicted in Fig.2 for some specific values of the considered network. 

\begin{figure}
  \includegraphics[width=\columnwidth]{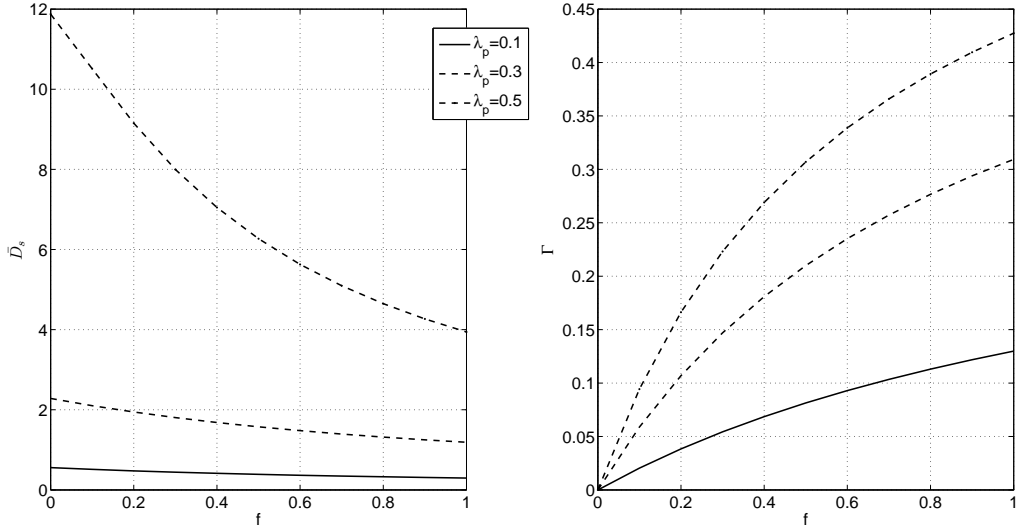}\\
  \caption{To relay or not to relay tradeoff based on the average delay $\bar{D_s}$ and power budget $\Gamma$}
  \label{fig:2}
\end{figure}

This behaviour led us to the definition of a minimization problem over the acceptance factor $f$ considering  the average delay $\bar{D_s}$  as an objective function subject to a constraint on the required power budget $\Gamma$. Taking into account the constraints imposed by keeping the stability of each queue, then the optimization problem can be defined as follows

\begin{eqnarray}\label{eq21}
\min_f & & \ \bar{D_s}(f)\\
&s.t.& \
\begin{cases}
\lambda_{p}\  < \mu_{p} \nonumber \\
\lambda_{ps} < \mu_{ps} \nonumber \\
\lambda_{s}\ < \mu_{s} \nonumber \\
\Gamma \ \ \leq \Gamma_{th}
\end{cases}
\end{eqnarray}
This problem is highly complicated and thus difficult to solve in this form due to the following reasons:
\begin{enumerate}
\item The objective function $\bar{D_s}$ is not easily differentiable over the acceptance factor  $f$
\item There are four complicated coupling constraints related to the acceptance factor $f$
\item The expectation is taking over three random variables $g_{ss}$,$g_{sp}$ and$g_{ps}$      
\end{enumerate}
Henceforth, we will adopt the concept of decomposition methods for solving this optimization problem ~\cite{c15}. We will use decomposition methods in order to decompose the original complicated problem into equivalent sub-problems as presented in ‎‎~\cite{c1}~\cite{c15}. The existing decomposition techniques can be classified into primal decomposition and dual decomposition methods ‎‎~\cite{c15}. The primal decomposition provides the decomposition of the original primal problem and it is appropriate when the objective function is formulated over two parameters. On the contrary, the dual decomposition is based on decomposing the Lagrangian dual problem derived from Lagrange multiplier application ~\cite{c15}. The problem is convex as it is proved in the Appendix although using decomposition the dual problem can always be considered convex.  

In our problem, we will follow an hierarchical decomposition that applies to our layered architecture by changing the primal and dual decompositions recursively. In particular, the basic decompositions are repeatedly applied to the problem to obtain smaller and smaller sub-problems. This approach corresponds to first applying a full dual decomposition, and then a primal one on the dual problem. As illustrated in Fig.3 the master problem is decomposed into two levels using dual decomposition for the two pairs of constraints where the first level includes one lower and one higher master problems and the second level includes the dual and primal decomposition of the resulted master problems. 

\begin{figure}
  \includegraphics[width=\columnwidth]{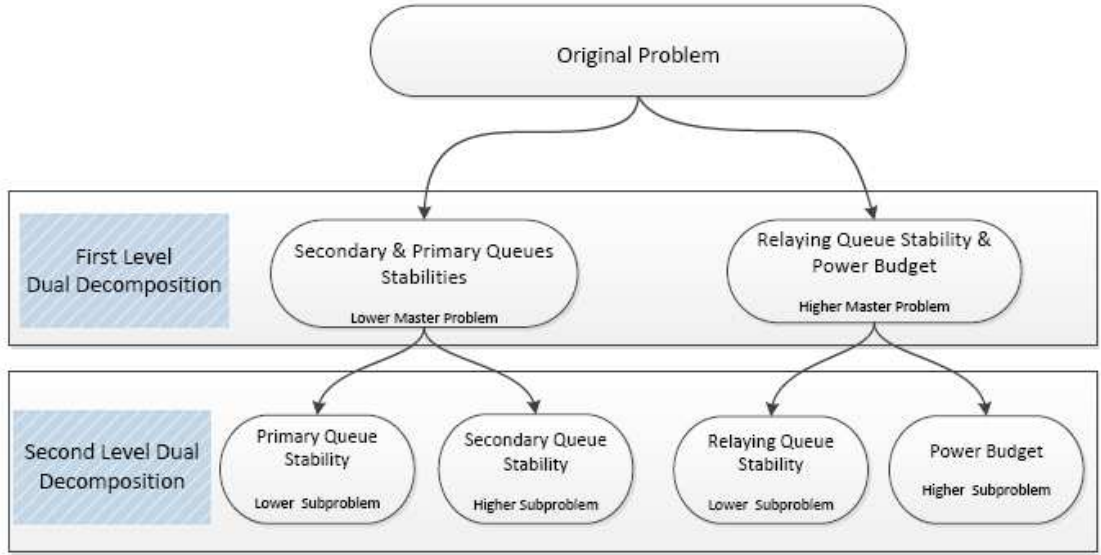}\\
  \caption{Hierarchical decomposition method}
  \label{fig:3}
\end{figure}

\subsection{First level decomposition}
For the problem in (\ref{eq21}), we will use first dual decomposition using the following two pairs of constraints. Which pair of constraints will be used is a matter of discussion and depends on how the system should be converged and stabilized. In general, using decomposition methods, the convergence and stability are guaranteed if the lower level problem is solved on a faster time scale than the higher level problem. Thus, we choose to have at the low level the stabilities of primary and secondary queues $Q_p$ and $Q_s$ since we want to converge first in these constraints and at the high level the stability of secondary-primary queue $Q_{ps}$ and power budget $\Gamma$ is retained that are more significantly demanding for the considered optimization problem.  

Going into details, while working at the first level, we obtain the Lagrangian function of (\ref{eq21}) considering the constraints of the stability of secondary-primary queue $Q_{ps}$ and power budget $\Gamma$ that can be written as follows
\begin{eqnarray} \label{eq22}
L_1 (f,\nu_1,\nu_2) = \bar{D_s} + \nu_1 (\lambda_{ps}-\mu_{ps}) + \nu_2(\Gamma-\Gamma_{th})
\end{eqnarray}
in which the Lagrangian dual problem is based and obtained as follows
\begin{eqnarray}\label{eq23}
\min_f & & \ \bar{D_s} + \nu_1 (\lambda_{ps}-\mu_{ps}) + \nu_2(\Gamma-\Gamma_{th}) \\
&s.t.& \
\begin{cases}
\lambda_{p}\  < \mu_{p} \nonumber \\
\lambda_{s}\ < \mu_{s} \nonumber \\
\end{cases}
\end{eqnarray}
The dual function in (\ref{eq22}) serves as a lower bound on the optimal value of the primal problem (\ref{eq20}). For instance, if we denote the minimization problem in (\ref{eq23}) as $g_1(\nu_1,\nu_2)$ and the optimal value of (\ref{eq21}) by $d_s^*$, then the inequality $d_s^* \geq g_1(\nu_1,\nu_2)$ holds for any non-negative values of $\nu_1$ and $\nu_2$. Having defined the lower master problem as in (\ref{eq23}) then the higher master problem using the dual optimization problem of (\ref{eq21}) is formulated  as

\begin{eqnarray}\label{eq24}
\max_{\nu_1,\nu_2} & & \ g_1(\nu_1,\nu_2) \\
&s.t.& \
\begin{cases}
\ \nu_1 \geq 0 \nonumber \\
\ \nu_2 \geq 0 \nonumber \\
\end{cases}
\end{eqnarray}

This approach provides in fact the solution of the dual problem instead of the original primal one and it will only give appropriate results if strong duality holds. The  problem in (\ref{eq24}) is convex even if the original problem is not convex because it is the point wise maximum of a family of affine function $(\nu_1,\nu_2)$, and thus the Karush–Kuhn–Tucker (KKT) conditions are satisfied and the duality gap is indeed zero ‎~\cite{c16}. 
Afterwards, we can solve the primal problem by solving the dual problem based on the dual function $g_1(\nu_1,\nu_2)$. This type of problem is solved by using iterative gradient or sub-gradient algorithms according to the differentiable capabilities of the dual problem. In our case, since the dual function is not differentiable, we will choose the sub-gradient solution ‎~\cite{c18}. The iterative algorithm will be described at the end of this section in conjunction with the solution of second level decomposition described in the next section.    

\subsection{Second level decomposition }
After the dual decomposition performed at the first level, the derived lower and higher master problems expressed by (\ref{eq23}) and (\ref{eq24}) respectively will be solved using dual and primal decomposition methods respectively. More specifically, the lower master problem presented by (\ref{eq23}) will be solved using dual decomposition since the optimization parameter is still the acceptance factor $f$. Thus, we obtain the Lagrangian function of (\ref{eq23}) considering the constraints of the stability of secondary and primary queues  that can be written as follows
\begin{eqnarray} \label{eq25}
\nonumber
L_2 (f,\xi) = \Bar{D_s} + \nu_1 (\lambda_{ps}-\mu_{ps}) \\
+ \nu_2(\Gamma-\Gamma_{th}) + \xi (\lambda_s-\mu_s)
\end{eqnarray}
in which the Lagrangian dual problem is based and obtained as follows
\begin{eqnarray}\label{eq26}
\min_f \ \bar{D_s} +  \nu_1 (\lambda_{ps}-\mu_{ps}) + \nu_2(\Gamma-\Gamma_{th}) + \xi (\lambda_s-\mu_s)\\
s.t. \
\begin{cases}
\lambda_{p}\  < \mu_{p} \nonumber \\
\end{cases}
\end{eqnarray}
For the known reasons discussed previously for the first level decomposition, the dual optimization problem of (\ref{eq23}) is written as
\begin{eqnarray}\label{eq27}
\max_{\xi \geq 0} \ g_2(\xi) 
\end{eqnarray}
where $g_2(\xi)$ is the minimization problem in (\ref{eq26}). Again, as we will see below the dual problem in (\ref{eq27}) will be solved by using an iterative sub-gradient algorithm. 

The higher master problem at the first level expressed by (\ref{eq24}) will be solved by using primal decomposition since the optimization is over two parameters i.e. $\nu_1$ and $\nu_2$. This will be separated again into two levels of optimization for each optimization parameter and thus two sub-problems are produced named lower and higher sub-problems. In the same notion, which sub-problem will be considered lower or higher depends on the convergence conditions that we are willing to retain. To this end, we prefer to keep at the higher level the optimization over the Lagrange multiplier $\nu_2$ that is related to the power budget constraint. Afterwards, the lower sub-problem is defined as follows over the Lagrange multiplier $\nu_1$   
\begin{eqnarray}\label{eq28}
\max_{\nu_1} \ g_1(\nu_1,\nu_2)\\
s.t. \
\begin{cases}
\nu_1 \  \geq 0 \nonumber \\
\end{cases}
\end{eqnarray}
while in parallel at the higher level, we have the following problem in charge of updating the coupling variable $\nu_2$  by solving 
\begin{eqnarray}\label{eq29}
\max_{\nu_2} \ g_1^*(\nu_1,\nu_2)\\
s.t. \
\begin{cases}
\nu_2 \  \geq 0 \nonumber \\
\end{cases}
\end{eqnarray}
where $g_1^*(\nu_1,\nu_2)$ is the optimal objective value of problem  (\ref{eq28})  for a given $\nu_2$. Both lower and higher sub-problems in  (\ref{eq28}) and  (\ref{eq29}) are convex since the original problem in  (\ref{eq24}) is a convex optimization as represents the dual problem of the master primal problem. Therefore, based again on sub-gradient method, we will use an iterative algorithm where the lower problem in  (\ref{eq28})  will converge first for given $\nu_2$. Details will be given below for the overall algorithm. 

If we gather all aforementioned decompositions in one formula, we will realize that the equivalent optimization problem of the original problem in (\ref{eq21}) can be written as follows
\begin{eqnarray}\label{eq30}
\nonumber 
\max_{\nu_2 \geq 0} \max_{\nu_1 \geq 0} \max_{\xi \geq 0} \min_f \ \bar{D_s} + \nu_1 (\lambda_{ps}-\mu_{ps}) \\
+ \nu_2(\Gamma-\Gamma_{th}) + \xi (\lambda_s-\mu_s) \\ 
s.t. \
\begin{cases}
\nonumber
\lambda_p \ < \mu_p \\
\end{cases}
\end{eqnarray}

Equation (\ref{eq30}) substitutes the original optimization problem in (\ref{eq21}) where the constraints have been introduced multiplied with the Lagrange multipliers  $\nu_1$,$\nu_2$  and $\xi$ that are actually the Karush–Kuhn–Tucker (KKT) conditions and they can be considered as shadow prices giving a freedom to potentially improve our objective function i.e. minimize the average delay. The additional decrease in the objective function is accomplished due to the relaxation of some of the given constraints ‎~\cite{c19}. Usually, how much performance improvement we can achieve by over-optimizing the objective function is determined by an allowable gap. However, as we will see from the simulation results, we can achieve better relaying tradeoffs without breaking the initial rules i.e. the constraints. All this discussion will throw more light on the formulated relaying tradeoff by looking into the sensitivity (shadow prices) analysis based on the formulated KKT conditions that retains the restrictions of the initial constraints ‎~\cite{c20}. Finally, the equivalent problem in (\ref{eq30}) will be solved by the following iterative subgradient-based algorithm as has been derived from the hierarchical decomposition described above. 

\begin{algorithm}
 \caption{Hierarchical Decomposition}
\begin{enumerate}
\item Initialization: $\nu_2^k$ and $k=1$ iteration
\item Repeat the initial values
\begin{enumerate}
\item Initialization: $\xi^j$ and $j=1$ iteration
\item Repeat the initial values
\begin{enumerate}
\item Calculate the acceptance factor $f$ from (\ref{eq26})
\item Calculate the sub-gradient at $\xi^j(\lambda_s-\mu_s)$  
\item Update $\xi^{j+1}$  by $\xi^{j+1}=\xi^j+\alpha \xi^j(\lambda_s-\mu_s)$, where $\alpha$ is the step size     
\end{enumerate}
\item Stop once $\mid \xi^{j+1}-\xi^j \mid \leq \varepsilon$ , where $\varepsilon$ is the convergence rule
\item Calculate the parameter $\nu_1^*$ using (29) based on the previous values $\xi$ and $f$  
\item Update $\nu_2^{k+1}$ by $\nu_2^{k+1}=\nu_2^{k}+\alpha(\nu_1^*)$, where $\alpha$ is the step size     
\end{enumerate}
\item Stop once $\mid \nu_2^{k+1}-\nu_2^k \mid \leq \varepsilon$ , where $\varepsilon$ is the convergence rule  
\end{enumerate}
 \end{algorithm}

The algorithm above is based on the subgradients of $g_1(\nu_1,\nu_2)$ and $g_2(\xi)$ given by the following propositions.

\textit{Proposition 1}: The subgradient of $g_1(\nu_1,\nu_2)$ is $s_1(f)=\mu_s-\lambda_s$ for the $j-th$ iteration and $s_1(f)_{\nu_1,\nu_2}$ is an element of $\partial s_1(\nu_1,\nu_2)$ \footnote[1]{$\partial s_1(\nu_1,\nu_2)$ denotes the set of all subgradients at $\nu_1$ and $\nu_2$ that is called the subdifferential.}

\textit{Proof}: For any $(\mu_1,\mu_2)\in dom(g_1)$ , since $g_1(\mu_1,\mu_2)$ is obtained by maximizing $L_1(f,\mu_1,\mu_2)$ over  $f\in dom(\bar(D)_s)$, we have $g_1(\mu_1,\mu_2)\geq L_1({f}_{\nu_1,\nu_2},\mu_1,\mu_2)$ \footnote[2]{where ${f}_{\nu_1,\nu_2}$ is the value of the acceptance factor $f$ for the Lagrange multiplier values $\nu_1$  and $\nu_2$.} ‎~\cite{c16}‎‎~\cite{c21}. Moreover, since ${f}_{\nu_1,\nu_2}$ achieves the maximum, we have $g_1(\nu_1,\nu_2)=L_1({f}_{\nu_1,\nu_2},\mu_1,\mu_2)$. Combining the pieces, we obtain

\begin{eqnarray} \label{eq31}
\nonumber
& & g_1(\mu_1,\mu_2) \geq L_1({f}_{\nu_1,\nu_2},\mu_1,\mu_2) \\
\nonumber 
&=& L_1({f}_{\nu_1,\nu_2},\nu_1,\nu_2) + L_1({f}_{\nu_1,\nu_2},\mu_1,\mu_2) - L_1({f}_{\nu_1,\nu_2},\nu_1,\nu_2)  \\
&=& g_1(\nu_1,\nu_2) + H({f}_{\nu_1,\nu_2})^T((\mu_1-\mu_2)-(\nu_1-\nu_2))
\end{eqnarray}     

Using a similar methodology, we can prove the existence of the subgradient $g_2(\xi)$ and then the problem can be solved by algorithm 1, which requires the calculation of the subgradients $g_1(\nu_1,\nu_2)$ and $g_2(\xi)$ at each of their own iteration.

\section{Relaying Behaviour for different $M/G/1$ Queuing Models }

One important issue in networks in general is to model the traffics of the primary users and secondary users ‎~\cite{c22}‎‎.  In this section, we will translate the acceptance factor $f$ into a specific metric of $M/G/1$ queuing model throwing more light on the performance impact in the considered cognitive radio model under realistic applications. More specifically, since the acceptance factor determines whether or not a new primary flow can be admitted to the secondary queue, we can merge it in the overflow probability and blocking probability for the infinite and finite buffer cases respectively ‎‎~\cite{c23}‎‎.   

\textit{Infinite buffer}: Although in case of infinite buffer, the notion of admission control can not be claimed, we can adopt the term of overflow or tail probability that is defined as the probability in which an arriving packet finds no room in the hypothetical queue. This aspect is identical to the example where in a waiting room, there is limited number of chairs for the customers, and any customer that upon arrival does not find a seat has to stand ~\cite{c24}. Let’s denote the limit of packets equal to $K$ and thus the overflow probability is expressed as follows 
	
\begin{eqnarray} \label{eq32}
p_{ov}=Pr[N\geq \lambda] = \sum_{n=K+1}^{\infty} p_n = 1-\sum_{n=0}^{K} p_n
\end{eqnarray}
where $N$ is the total number of packets in the system and $p_n$  is the queue state, namely the probability that there are   packets in the queue. The probability $p_n$ is obtained as the Laplace inverse transform of its generating function defined as follows
\begin{eqnarray} \label{eq33}
P_n(Z)= \frac{(1-\rho)\bar{B}(\lambda-\lambda z)(1-z)}{\bar{B}(\lambda-\lambda z)-z}
\end{eqnarray}
which represents the Pollaczek-Khinchin formula ~\cite{c25}, where the Laplace inverse transform when Poisson arrival process and  generalized service time are assumed results in the following expression for the probability $p_n$ 
\begin{eqnarray} \label{eq34}
p_n= \rho + \frac{\lambda^2 E[S^2]}{2(1-\rho)}
\end{eqnarray}
Substituting (\ref{eq34}) into (\ref{eq32}), we can derive the overflow probability for the infinite $M/G/1$ queue that is associated with the acceptance factor through the arrival rate and the utilization factor of the queues in general. 

\textit{Finite buffer}: In case of finite buffer the blocking probability plays the role of admission control metric in admission control policy. A well-known example of finite systems is the blocking probability in cellular networks where the limit resources can be fully allocated. There are numerous approximations for the blocking probability possible for $M/G/1/K$ queuing systems as described in ~\cite{c26}. A good approximation of the blocking probability $p_b$ is the one that is based on the overflow probability of an infinite buffer queue that we discuss above and thus we can associate it with the analysis of the overflow probability. To this end, the blocking probability is defined as follows
\begin{eqnarray} \label{eq35}
p_b= \frac{(1-\rho)p_{ov}}{1-\rho p_{ov}}
\end{eqnarray}
where $p_{ov}$ is the overflow probability obtained by (\ref{eq32}) using (\ref{eq34}) and eventually the blocking probability is associated with the acceptance factor as mentioned above. 

In order to solve the problem in (\ref{eq21}) over the overflow probability $p_{ov}$ defined in (\ref{eq32}) and the blocking probability $p_b$ defined in $\ref{eq35}$, we need to discuss for the convexity of these equations over the utilization factor $\rho$ that is a linear function of acceptance factor $f$ and thus we can infer that are convex over $f$ as well. 

\textit{Proposition 2}: For the range of value $\rho$, $0 < \rho < 1$  the  first derivative of $p_{ov}(\rho)$ is increasing and convex on $\rho$.  

\textit{Proof:}  Taking the first derivative of overflow probability $p_{ov}$ in (13) with respect to $\rho$ gives:

\begin{eqnarray} \label{eq36}
p_{ov}^{'}(\rho) = \frac{\lambda^2E[S^2]}{(\rho-1)^2}+1
\end{eqnarray}

For $0 < \rho < 1$, it is evident that $p_ov^{'}(\rho)>0$ and thus $p_{ov}(\rho)$ is increasing in $\rho$ i.e. is convex in $\rho$. 

Regarding the convexity of blocking probability in (\ref{eq32}), it can be considered as $(1-\rho)x/(1-\rho x)$ that is convex increasing to the interval $\rho\in[0,1]$ with $f^{'}(0)=(1-\rho)$ and $f(1)=1$ as discussed in ~\cite{c27}.  

\section{Simulation Results}

We provide below the simulation results of the optimization problem above with the corresponding discussion. We assume the following values for the SNR thresholds $\gamma_{th,p}=\gamma_{th,ps}=\gamma_{th,sp}=\gamma_{th,s}=0dB$  and the following channel variables $\sigma_p^2=4dB$,$\sigma_{ps}^2=12dB$,$\sigma_{sp}^2=8dB$  and $\sigma_{s}^2=12dB$. The transmission power needed for each queue is $P_p=1$, $P_s=1$  and $P_{sp}=0.25$ . Fig. 4a depicts the optimal acceptance factor $f^*$ and the corresponding average delay $\bar{D_s}(f^*)$ vs. arrival rate at the primary Queue $\lambda_p$  for $\Gamma_{th}=0.2$ power budget threshold and $\lambda_s=0.1$ as secondary arrival rate. The figure gives insights of how much relaying is needed for different values of the primary loading. For small values of  $\lambda_p$, there  is low traffic in the primary queue, and the SU-Tx (i.e. cognitive node) has more opportunities to access the channel, so it can accept all the undelivered primary packets for relaying while wasting a small percentage of its power. As increasing the primary load increases the number of relayed packets, the cognitive node reduces the accepted fraction of these packets to keep the predefined power budget. For high values of $\lambda_p$, the SU-Tx might choose not to relay the primary packets, as a small reduction in the secondary delay is associate with very high power consumption. In the same notion, Fig. 4b depicts the optimal acceptance factor $f^*$ and the corresponding average delay $\bar{D_s}(f^*)$   vs. arrival rate at the secondary Queue $\lambda_s$ for $\Gamma_{th}=0.2$  power budget threshold and $\lambda_p=0.3$ as secondary arrival rate. The insight is opposite of the previous one i.e. vs. arrival rate at the primary queue $\lambda_p$. For high values of $\lambda_s$  the cognitive  i.e. secondary node does not has enough opportunities to access the channel since the primary packet are mainly forwarded as pointed out above. 

\begin{figure}
  \includegraphics[width=\columnwidth]{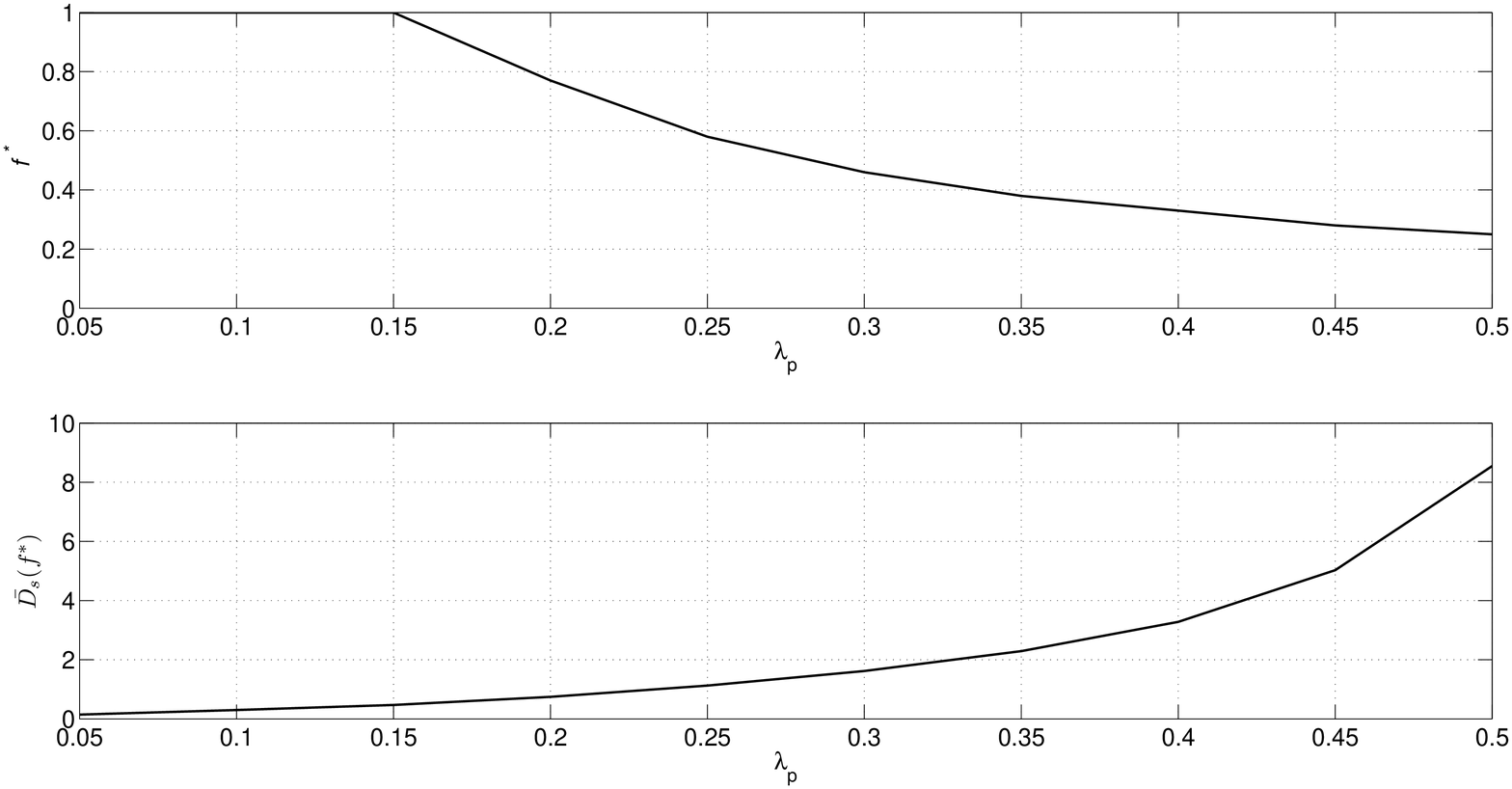}\\
  \caption{Optimal acceptance factor $f^*$ and corresponding average delay $\bar{D_s}(f^*)$  vs. arrival rate at the primary Queue $\lambda_p$}
  \label{fig:4a}
\end{figure}

\begin{figure}
  \includegraphics[width=\columnwidth]{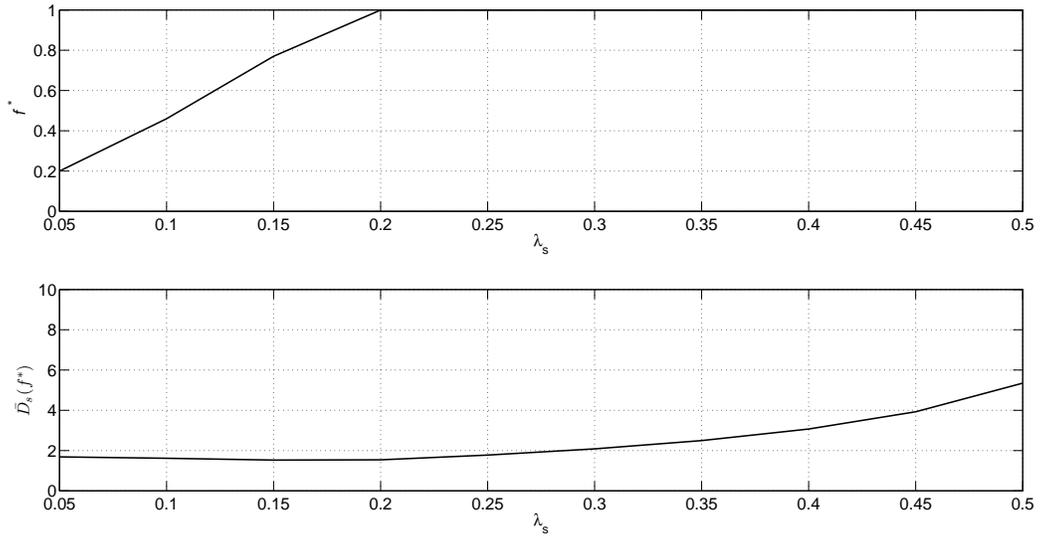}\\
  \caption{Optimal acceptance factor $f^*$ and corresponding average delay $\bar{D_s}(f^*)$  vs. arrival rate at the secondary Queue $\lambda_s$}
  \label{fig:4b}
\end{figure}

Fig. 4c depicts the optimal acceptance factor $f^*$ and corresponding average delay $\bar{D_s}(f^*)$  vs. power budget threshold $\Gamma_{th}$. The effect of the relaying power budget threshold $\Gamma_{th}$ on the secondary delay $\bar{D_s}$  is shown. The solution for the optimization problem for different values of $\Gamma_{th}$ is plotted. The simulation parameters are the same as that for the previous figures and for arrival rates we have obtained $\lambda_p=0.5$ and $\lambda_s=0.1$. The general interpretation of this results is that if low power budget is assigned to the relaying channel (i.e. $ 0 \leq \Gamma_{th} \leq 1 $) , the cognitive sensor node accepts a little fraction of the undelivered primary packets ($f\leq 0.1$) which does not result in a significant reduction in the secondary delay. However, increasing the budget threshold allows more packets to be accepted for relaying and, in turn, reduces the secondary delay but with a limit. The figure shows that, spending more than $0.45$ of the total power on relaying, will not result in any reduction in the secondary node delay as it accepts all the undelivered primary packets for relaying.

\begin{figure}
  \includegraphics[width=\columnwidth]{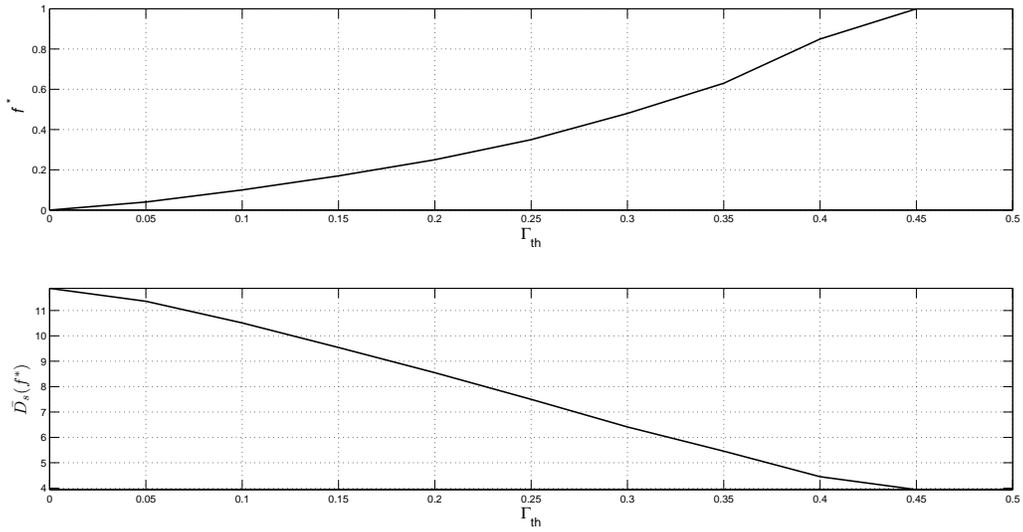}\\
  \caption{Optimal acceptance factor $f^*$ and corresponding average delay $\bar{D_s}(f^*)$ vs. power budget threshold  $\Gamma_{th}$}
  \label{fig:4c}
\end{figure}

Fig. 5a depicts the sensitivity of the average secondary delay $\bar{D_s}$ in terms of the KKT conditions of the original optimization problem in (\ref{eq21})  as they introduced by the dual problem in (\ref{eq30}). We keep the same values for the parameters as previously. In three axes the KKT conditions $\nu_1(\lambda_{ps}-\mu_{ps})$, $\nu_2(\Gamma-\Gamma_{th})$ and $\xi(\lambda_s-\mu_s)$ are denoted as $\nu_1(f)$, $\nu_2(f)$ and $\xi(f)$ respectively at x, y and z axes as well. Hence, they can be called shadow prices since they can provide decision-making with powerful insights into our problem. For instance, it is obvious from the Fig.5a that the KKT condition related to the power budget threshold i.e. $\nu_2(f)$ might not be kept as long as the Lagrange multiplier $\nu_2$  is increased more than a threshold since obviously the  $\nu_2(f)$  is increased in this case beyond the zero value. This is applied when the primary arrival rate $\lambda_p$ is getting higher i.e $\lambda_p>0.3$. In the contrary, the more increase of the Lagrange multipliers $\nu_2$  and $\xi$, the more the decrease of the KKT conditions $\nu_1(f)$ and $\xi(f)$ is become and as a consequence the average delay $\bar{D_s}$  of the dual problem in (\ref{eq30}). In Fig.5b, we depict the KKT conditions changing the secondary arrival rate $\lambda_s$ keeping the primary arrival rate $\lambda_p=0.1$  and we highlight that all shadow prices $\nu_1(f)$,$\nu_2(f)$  and $\xi(f)$  can potentially decrease the average delay without having a harmful impact on the stability of secondary queue.  

\begin{figure}
  \includegraphics[width=\columnwidth]{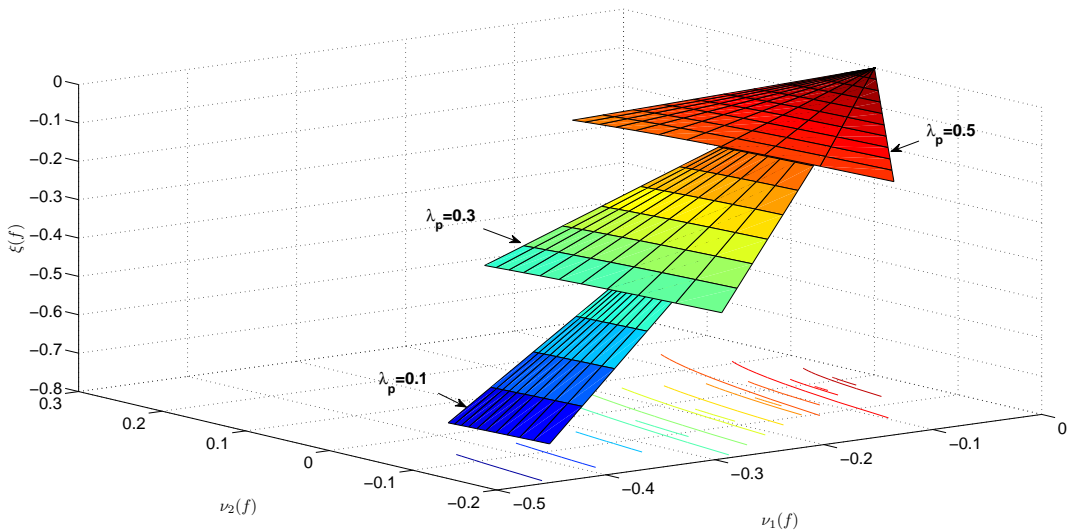}\\
  \caption{ Arrival’s rate $\lambda_p$ sensitivity of primary queue $Q_p$ on shadow prices i.e. Lagrange multipliers}
  \label{fig:5a}
\end{figure}

\begin{figure}
  \includegraphics[width=\columnwidth]{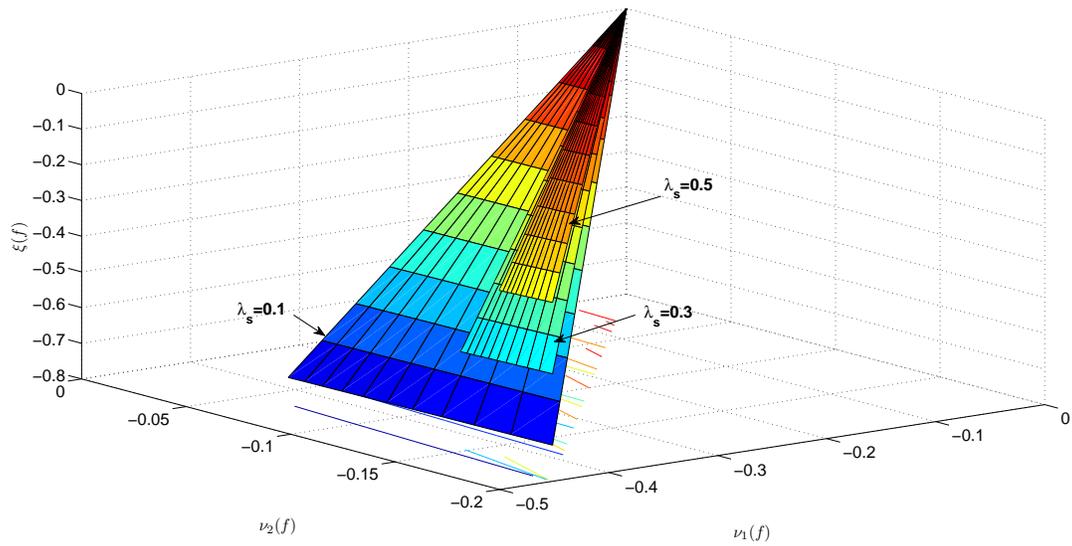}\\
  \caption{ Arrival’s rate $\lambda_s$ sensitivity of secondary queue $Q_s$ on shadow prices i.e. Lagrange multipliers}
  \label{fig:5b}
\end{figure}

Fig. 5c depicts the sensitivity of the power’s budget $\Gamma$ sensitivity on shadow prices i.e. Lagrange multipliers for different threshold values $\Gamma_{th}$. We retain the aforementioned parameters for the CRSN and for the arrival rates we assume $\lambda_s=0.1$ and $\lambda_p=0.5$. Notably, as long as the power budget threshold $\Gamma_{th}$ is low, the increase in $\nu_2(\Gamma-\Gamma_{th})$  KKT condition will not pass the intolerable limit of increasing the average delay. However, the decrease is retained for low values of the corresponding shadow price $\nu_2(f)$ and not for the high power budget threshold values i.e. $\Gamma_{th}=0.4$ and $\Gamma_{th}=0.6$.  

\begin{figure}
  \includegraphics[width=\columnwidth]{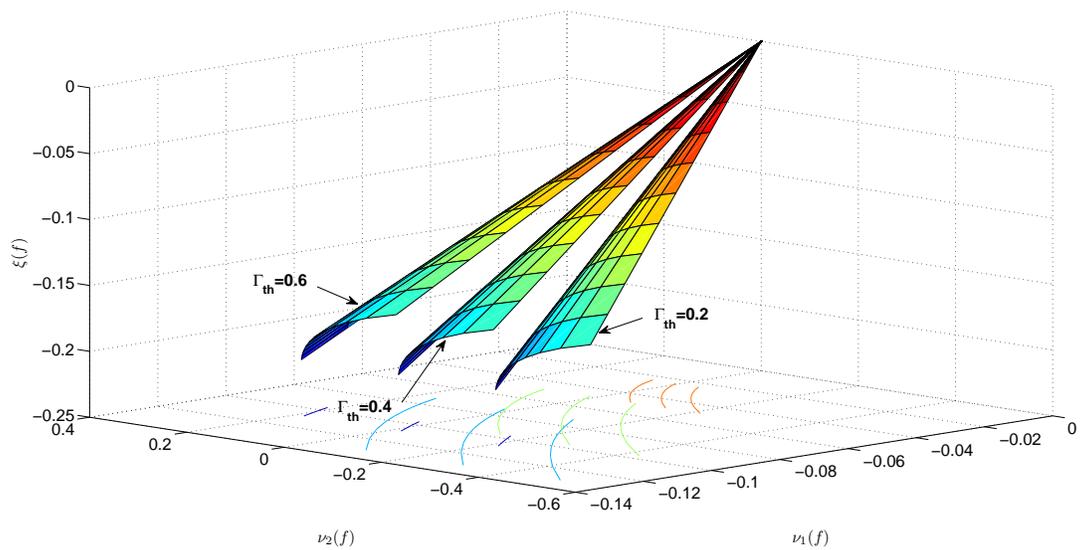}\\
  \caption{Power’s budget $\Gamma$ sensitivity on shadow prices i.e. Lagrange multipliers}
  \label{fig:5c}
\end{figure}

Fig. 6 depicts in the top axis the overflow and blocking probabilities denoted as $p_{ov}$ and $p_b$ respectively for different values of primary arrival rate $\lambda_p$. We assume that the secondary arrival rate is $\lambda_s=0.1$ and the threshold on the power’s budget $\Gamma_{th}=0.2$. In the same concept, the middle axis depicts the overflow and blocking probabilities $p_{ov}$ and $p_b$ respectively for different values of secondary arrival rate $\lambda_s$. We assume that the primary arrival rate is $\lambda_p=0.3$ and the threshold on the power’s budget $\Gamma_{th}=0.2$. Finally, the bottom axis depicts the overflow and blocking probabilities $p_{ov}$ and $p_b$ respectively for different values of threshold on the power’s budget $\Gamma_{th}$ with $\lambda_p=0.5$ and $\lambda_s=0.1$. From these three figures is inferred that the overflow probability is always more than the blocking probability in general and this is due to the relaxing behaviour of infinite buffers against the blocking probability of finite buffers. On the other hand, the convex properties of both probabilities are confirmed as it is highlighted in ~\cite{c26}. More remarks on the behaviour of relaying queue are that first the overflow and blocking probabilities are not increased up to one level over secondary arrival rate  due to the forwarding mainly of the primary packets process as pointed out above and finally the probabilities tends to one as long as the power budget threshold is relaxed. 

\begin{figure}
  \includegraphics[width=\columnwidth]{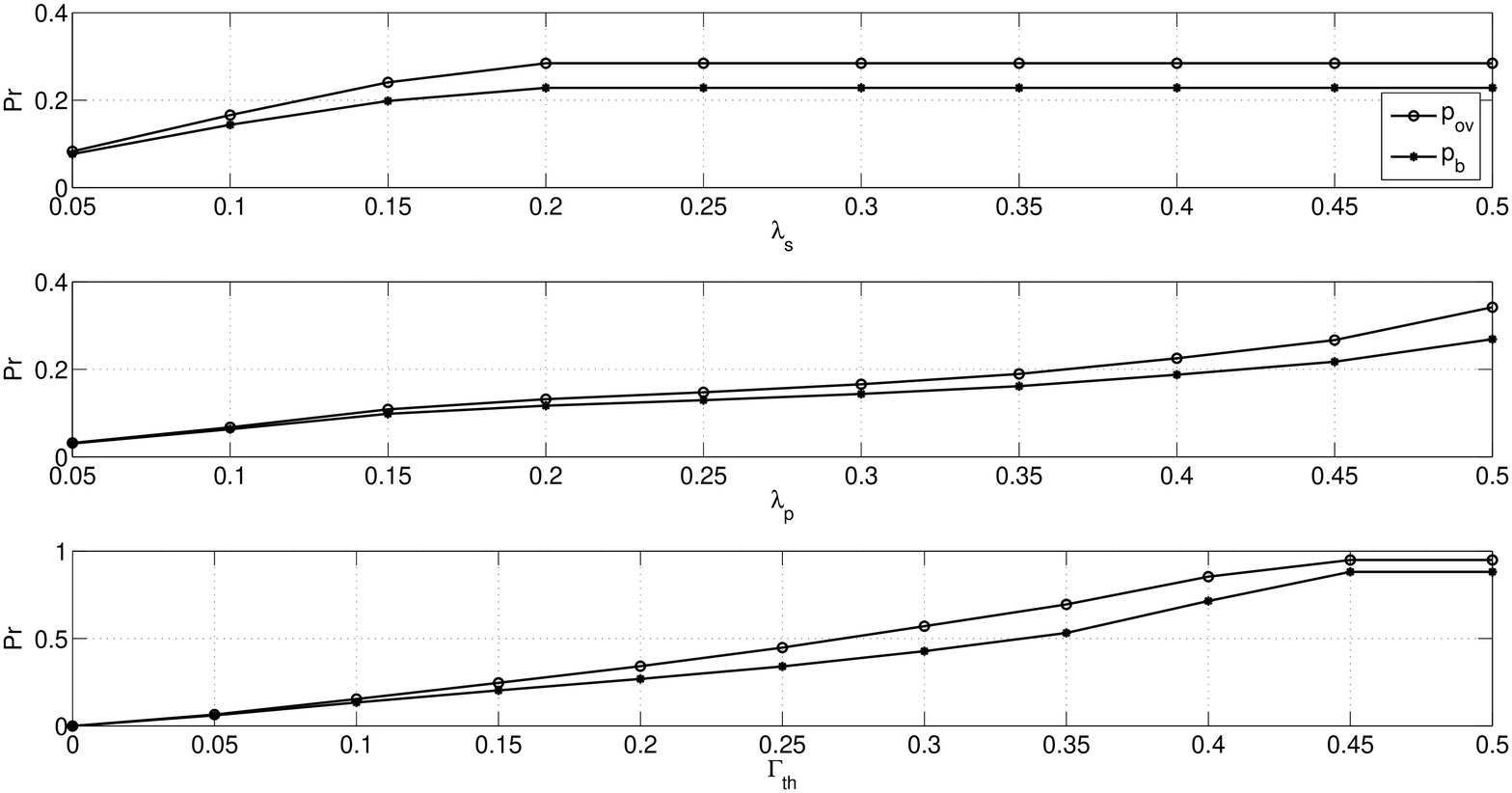}\\
  \caption{Overflow and blocking probabilities for the optimal acceptance factor vs. primary arrival rate $\lambda_p$, secondary arrival rate $\lambda_s$ and power budget threshold $\Gamma_{th}$ respectively}
  \label{fig:6}
\end{figure}

\section{Summary}
In this paper, we consider a sensor system in which a node is able to relay the primary packets at MAC layer when the primary cannot achieve adequate communication. The relaying model is based on queue modelling where the stability rules should be satisfied accompanying with the customized rules of our investigation, which are the minimum delay induced in the secondary packets and the power budget required for relaying the primary packets. The relaying capability is controlled by an admission control factor that is associated with all available distributed queues in the system. Based on the considered model, we define an optimization problem with four several constraints. Specifically, the problem is formulated as the minimization of average delay at secondary packets as long as the queues’ stabilities and the power limit are guaranteed. We solve this complex optimization problem using hierarchical decomposition method thereby the several constraints are imposed separately applied first and second level decomposition, which finally are broken into more concrete sub-problems. This method results in the equivalent problem that encompass all the constraint requirements as KKT conditions. We obtain results that highlight useful outcomes in terms of system performance as well as indicate the sensitivity strength of  the objective function both to the separate numerous constraints. We also assess our solution considering infinite and finite buffer capacity for the queue that forwards the primary packets. 

\appendix
\section{Convexity of $\bar{D}_s(f)$ over $f$}
\textit{Proposition 3}: For the range of value $f$, $0\leq f \leq 1$  the  first derivative of $\bar{D}_s(f)$ is negative and increasing (i.e. $\bar{D}_s(f)\leq 0$) and thus the $\bar{D}_s(f)$ is decreasing and convex on $f$.  

\textit{Proof: } Looking into the average secondary delay $\bar{D}_s(f)$  in (13), both  $\rho_p$ and $\rho_s$  utilization factors are constants while the $\rho_{ps}$ and the average primary delay $\bar{D_p}$ do not. Hence, differentiating $\bar{D}_s(f)$  with respect to $f$ gives:

\begin{eqnarray} \label{eq37}
\bar{D}_s^{'}(f) = A \rho_{ps}^{'}(f) + B \bar{D_p}^{'}(f) + C
\end{eqnarray}

where $A$, $B$ and $C$ are some constants while $\rho_{ps}^{'}(f)$  and $\bar{D_p}^{'}(f)$  are the first derivatives of utilization factor and primary delay respectively which are obtained as follows
\begin{eqnarray} \label{eq38}
\bar{D_p}^{'}(f) = \frac{(1-\lambda_p)P_{out,p}(1-P_{out,ps})}{(1-\lambda_p-P_{out,p}+f P_{out,p}(1-P_{out,ps}))^2}
\end{eqnarray}
and 
\begin{eqnarray} \label{eq39}
\nonumber
& & \rho^{'}_{ps}(f) = \frac{1-\lambda_p}{(1-P_{out,p})+f P_{out,p}(1-P_{out,ps})-\lambda_p}\\
&-& \frac{f \lambda_p P_{out,p}^2(1-P_{out,ps})^2}{(1-P_{out,p}+f P_{out,p}(1-P_{out,ps}))^2}
\end{eqnarray}
For $0\leq f \leq 1$  it is clear that $\bar{D_s}^{'}(f)<0$ i.e. negative and increasing as well and thus $\bar{D_s}(f)$ is decreasing as depicted in Fig.2 and convex as a conclusion.

\end{document}